\documentclass[%
 aip,
 amsmath,amssymb,
 reprint,%
]{revtex4-1}

\usepackage{graphicx}
\usepackage{dcolumn}
\usepackage{bm}
\usepackage{siunitx}
\usepackage{textcomp}
\usepackage[utf8]{inputenc}
\usepackage[T1]{fontenc}
\usepackage{mathptmx}
\usepackage{xcolor}
\usepackage{gensymb}
\usepackage[english]{babel}
\begin{document}

\preprint{AIP/123-QED}
\title[]{Image-driven discriminative and generative machine learning algorithms for establishing microstructure-processing relationships}
\author{W. Ma}%
\affiliation{Computer Science Department, Rensselaer Polytechnic Institute}

\author{E.J. Kautz}
 \email{elizabeth.kautz@pnnl.gov}
 \affiliation{National Security Directorate, Pacific Northwest National Laboratory}

%

\author{A. Baskaran}
\affiliation{%
Materials Science and Engineering Department, Rensselaer Polytechnic Institute
}%

\author{A. Chowdhury}
\affiliation{%
Artificial Intelligence, GE Research
}%


\author{V. Joshi}
\affiliation{%
Energy and Environment Directorate, Pacific Northwest National Laboratory
}%

\author{B. Yener}
\affiliation{%
Computer Science Department, Rensselaer Polytechnic Institute
}%

\author{D.J. Lewis}
\email{lewisd2@rpi.edu}
\affiliation{Materials Science and Engineering Department, Rensselaer Polytechnic Institute
}%

\date{\today}

\begin{abstract}
We investigate methods of microstructure representation for the purpose of predicting processing condition from microstructure image data. A binary alloy (uranium-molybdenum) that is currently under development as a nuclear fuel was studied for the purpose of developing an improved machine learning approach to image recognition, characterization, and building predictive capabilities linking microstructure to processing conditions. Here, we test different microstructure representations and evaluate model performance based on the F1 score. A F1 score of 95.1\% was achieved for distinguishing between micrographs corresponding to ten different thermo-mechanical material processing conditions.  We find that our newly developed microstructure representation describes image data well, and the traditional approach of utilizing area fractions of different phases is insufficient for distinguishing between multiple classes using a relatively small, imbalanced original data set of 272 images. To explore the applicability of generative methods for supplementing such limited data sets, generative adversarial networks were trained to generate artificial microstructure images. Two different generative networks were trained and tested to assess performance. Challenges and best practices associated with applying machine learning to limited microstructure image data sets is also discussed. Our work has implications for quantitative microstructure analysis, and development of microstructure-processing relationships in limited data sets typical of metallurgical process design studies.
\end{abstract}

\maketitle

\section{\label{sec:level1}Introduction}
Microstructure image data is rich in information regarding morphology and implied composition of constituent phases, and can provide unique insight into the pathways leading to microstructure formation, and mechanisms responsible for material behavior and performance. Thus, the analysis of micrographs (i.e. microstructure image data) is central to several materials science studies establishing processing-structure-property relationships, and for the design of new material systems. Despite the ubiquity of micrographs in material science research, significant challenges exist related to consistent and accurate recognition and analysis of image data. Such challenges arise from the domain knowledge and skill required to obtain micrographs, the diverse types of image data possible (e.g. optical and electron microscopy), domain-specific challenges to image analysis techniques, and more. With the advancing application of artificial intelligence (AI) (i.e. machine learning) in a wide range of fields, we find the application of established AI methods to microstructure recognition and analysis opens up an opportunity for computationally-guided experiments, and objective, repeatable analysis of image data. To address the need for improved microstructure quantification via image-driven machine learning using small, imbalanced data sets, we investigate microstructure-processing relationships in a model binary uranium-molybdenum (U-Mo) alloy.  The U-Mo system is of particular interest due to the alloy's applicability as a nuclear fuel for research reactors, and the need to understand microstructure-processing relationships for improved fabrication design and fuel qualification.

Uranium (U) alloyed with 10 weight percent (wt\%) molybdenum (Mo), referred to here as U-10Mo, is currently under development as a new metallic nuclear fuel for application in research and radioisotope production facilities. U-10Mo is a candidate for low enriched U (LEU) fuel, designed to replace currently used highly enriched U (HEU) fuels with the aim of reducing proliferation and safety risks associated with HEU handling and operation\cite{NAS2016,Berghe2014, Snelgrove1997, Meyer2014, IAEA1996}. A monolithic, plate-type design for U-10Mo has been selected due to the high U densities achievable while meeting the low enrichment specification, where the fuel must have $\leq$ 20\% \textsuperscript{235}U relative to all U isotopes. In order to fabricate fuel plates to meet dimensional requirements, the U-10Mo alloy must be subjected to several thermo-mechanical processing steps, leading to microstructural evolution during fabrication (e.g. hot rolling, hot isostatic pressing). To design a fuel with microstructure that meets performance requirements, and to enable future materials processing design, the microstructure-processing relationship must be well-established.

The equilibrium phase of pure U at room temperature ($\alpha$-U) has an orthorhombic crystal structure. $\alpha$-U is known to experience non-uniform thermal expansion in a high temperature, irradiation environment, thus 10 wt\% Mo is added to stabilize the high-temperature BCC $\gamma$-U phase at room temperature.  During processing, U-10Mo is exposed to temperatures below 560-575$\degree$C during hot isostatic pressing (HIP). Below these temperatures, a eutectoid decomposition of the metastable $\gamma$-UMo matrix phase into $\alpha$-U and $\gamma'$ (U$_2$Mo) is expected, based on the equilibrium binary phase diagram \cite{UMoPhaseDiagram}. Prior work has demonstrated that this eutectoid decomposition occurs via a discontinuous precipitation (DP) mechanism. The decomposition involves the $\gamma$-UMo matrix phase transforming to $\alpha$-U and Mo-enriched $\gamma$-UMo products with lamellar morphology. Our previous work showed that this transformation was initiated primarily at grain boundaries, and interphase interfaces, where Si segregation was observed\cite{Devaraj2018_Acta, Devaraj2018_Scripta,Jana2017a,Kautz_SciRep}.

Significant prior work has been performed using machine learning for a range of materials science applications \cite{npjComp_Ramprasad,Ling2017,Chowdhury2016, DeCost2018,DeCost2015,DeCost2017,Butler2018,Kalinin2016, Ballard2017,Pilania2013_SciRep}.  A rapidly growing area in machine learning in materials science is in image data  quantification. Previous studies demonstrated success of convolutional neural networks (CNNs) in microstructure recognition tasks without significant development time, and state-of-the-art performance for a wide range of microstructures (e.g. forged titanium, perlitic steel, metal powder, ceramics) \cite{DeCost2015,DeCost2017,DeCost2018,Ling2017,Kondo2017,Baskaran2020, Chowdhury2016}. Additionally, the application of machine learning methods to large image data sets, such as those available through ImageNet is routinely done \cite{Krizhevsky2012,Deng2009}. The ImageNet database includes over 14 million natural images that can be used for training and testing of machine learning models. Application of machine learning methods to limited data sets is still a frontier in the machine learning community \cite{Butler2018, Lake2015,Kondo2017,Kautz2019,Mace2018}, and is of interest to materials science data analysis problems, where only limited, unbalanced, or historic data sets are available, and the cost/time associated with obtaining very large data sets is prohibitive.

The present study explores the applicability of image-driven machine learning methods \cite{Chowdhury2016} to developing microstructure-processing relationships. Specifically, we seek to understand the role of several thermomechanical processing steps on the microstructure evolution observed in the U-10Mo system. An improved approach to determining microstructure-processing relationships is developed and presented here, involving feature extraction, segmentation, and classification using a random forest model. Microstructure image data is segmented to identify microstructural features of interest and quantify area fraction of these features, including the $\gamma$-UMo matrix, uranium carbide, and DP reaction transformation products. The application of generative adversarial networks (GANs) \cite{10.5555/2969033.2969125, pan2019recent} is also discussed as an emerging method for microstructure image generation. Our work has broad implications for machine learning applications in microstructure image analysis, and the development of quantitative microstructure-processing relationships in a wide range of alloy systems.

\section{Experimental and Computational Methods}
\subsection{\label{sec:methods_2}Image Data}


Image data used in this work is from two scanning electron microscopes (SEMs): a FEI Quanta dual beam Focused Ion Beam/Scanning Electron Microscope (FIB/SEM), and a JEOL JSM-7600F Field Emission SEM. The backscatter electron detector was used for improved atomic number (Z) contrast. Two different microscope operators took the images. Thus, the image data analyzed here was diverse in terms of resolution, contrast, focus, and magnifications selected. The idea in using a variety of images taken by different operators using different microscopes (but all of the same samples) was to develop a more robust model that can distinguish between different material processing conditions.

\begin{figure*}
    \centering
    \includegraphics[scale=0.9]{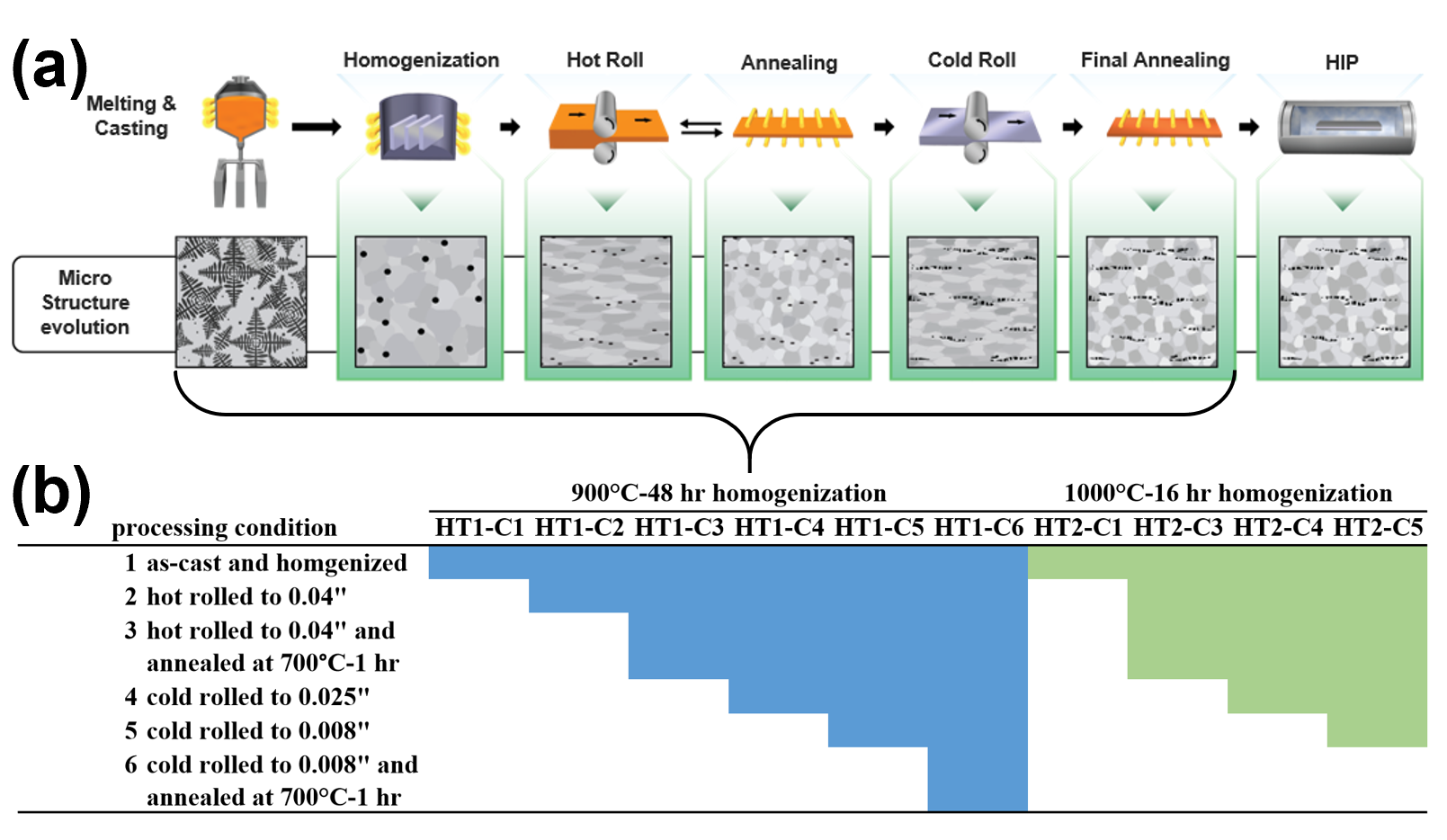}
    \caption{(a) Schematic of U-10Mo fuel fabrication, where the steps shown in the bracket were used to generate microstructure imaged and analyzed in this work. (b) Sample matrix of sample processing conditions analyzed. For each condition, SEM micrographs were analyzed. Two different homogenization temperatures were utilized, and thus samples are grouped based on the homogenization treatment performed. Sample conditions are indicated by homogenization treatment (HT) number, where HT-1 refers to 900\textdegree C-48hr, and HT-2 refers to 1000\textdegree C-16hr. The processing condition (C) the micrographs represent are indicated by C followed by a number that corresponds to the number processing condition (e.g. C1 is a sample in the as-cast condition).}
    \label{fig:sample_matrix}
\end{figure*}

All images used in this work are of a depleted U-10Mo alloy fabricated and prepared according to details presented elsewhere \cite{Jana2017, Prabhakaran2016}. Images were taken over a range of magnifications from 250x to 500x. Image data was labeled based on processing condition, detailed in Figure \ref{fig:sample_matrix}. Ten different image classes were studied, where each class corresponds to a different processing history that generates a unique microstructure. The processing conditions detailed in Figure \ref{fig:sample_matrix} include two different homogenization annealing treatments (900\textdegree C-48hr and 1000\textdegree C-16hr) and several thermo-mechanical processing steps such as cold and hot rolling. Each image class is therefore labeled by the homogenization treatment (HT) and processing condition (C) numbers, where 900\textdegree C-48hr is referred to as HT1, and 1000\textdegree C-16hr is HT2. Processing conditions are indicated by C followed by the number in the list of all possible conditions in Figure \ref{fig:sample_matrix}. For example, HT1-C1 is a U-10Mo sample that is in the as-cast and homogenized condition, where homogenization was done at 900\textdegree C for 48 hours.

Original images vary in size. Different image sizes used in this work (in pixels) include:  2048 by 2560, 1448 by 2048, 1428 by 2048, and 1024 by 1280. All images included a scale bar region which was removed prior to training and testing by cropping the image. The data set analyzed here consists of a total of 272 original images from 10 classes. Bilateral filters were applied to each image for noise removal while keeping edges sharp. In our study, we chose the diameter of each pixel neighborhood as 15 while keeping all other default parameters.

\subsection{Discriminative Methods}
\subsubsection{\label{sec:methods_2a}Feature Extraction}
In order to determine how to best quantify microstructure image data on the U-10Mo system (as a function of thermo-mechanical processing parameters), different methods of feature extraction were developed and tested. Here, each microstructure image is described by a feature vector, and how that feature vector is derived either depends on area fraction of different regions, or spatial relationships between microstructural features of interest. These two types of features are referred to here as area and spatial features, respectively. These two different feature types are extracted from each image after segmentation. Area features are simply the area fractions of each phase or region ($\gamma$-UMo matrix, UC, and lamellar transformation products). U-10Mo microstructures have been described by the area fractions of these regions in prior work.\cite{kautz2019imagedriven,Jana2017,Jana2017a} Spatial features are computed by first measuring the following for each region (matrix, carbide, lamellar transformation products): the $x$ and $y$ coordinate of the centroid, area (in square pixels), and the ratio of area of the region to the area of its bounding box. The spatial feature is simply a concatenation the following measures: the number of regions, the mean and standard deviation of the areas, the standard deviation of the centroid coordinates, and the mean and standard deviation of area ratios.

\subsection{\label{sec:methods_2} Approach and Machine Learning Model}
All experimentation was carried out with Python version 3.6.9 with the help  of various open-source libraries. The opencv, scipy, skimage, numpy, and sklearn packages (compatible with Python version 3.6.9) were used for training, testing, and validation. All relevant model parameters are summarized in Table \ref{tab:parameters}.

\begin{table}
\caption{\label{tab:parameters} Parameters selected for  model specification, compilation, and cross validation for U-10Mo image analysis.
 }
\begin{ruledtabular}
\begin{tabular}{|r|c|c|}
&\mbox{\textbf{Parameter}}&\mbox{\textbf{Value}}\\ \hline

\textbf{Preprocessing} & Noise-reducing &	Bilateral \\
    &	diameter	&	15	\\
	&	sigma color	&	75	\\
	&	sigma space	&	75	\\	\hline
\textbf{Segmentation} & k-means & kmeans++ center initialization \cite{10.5555/1283383.1283494} \\
	&	Lamellar	&	(9,9) closing then (9,9) opening	\\
	&	UC	&	(9,9) opening then (3,3) closing	\\ \hline
\end{tabular}
\end{ruledtabular}
\end{table}

\begin{figure}
    \centering
    \includegraphics[height=2.4in]{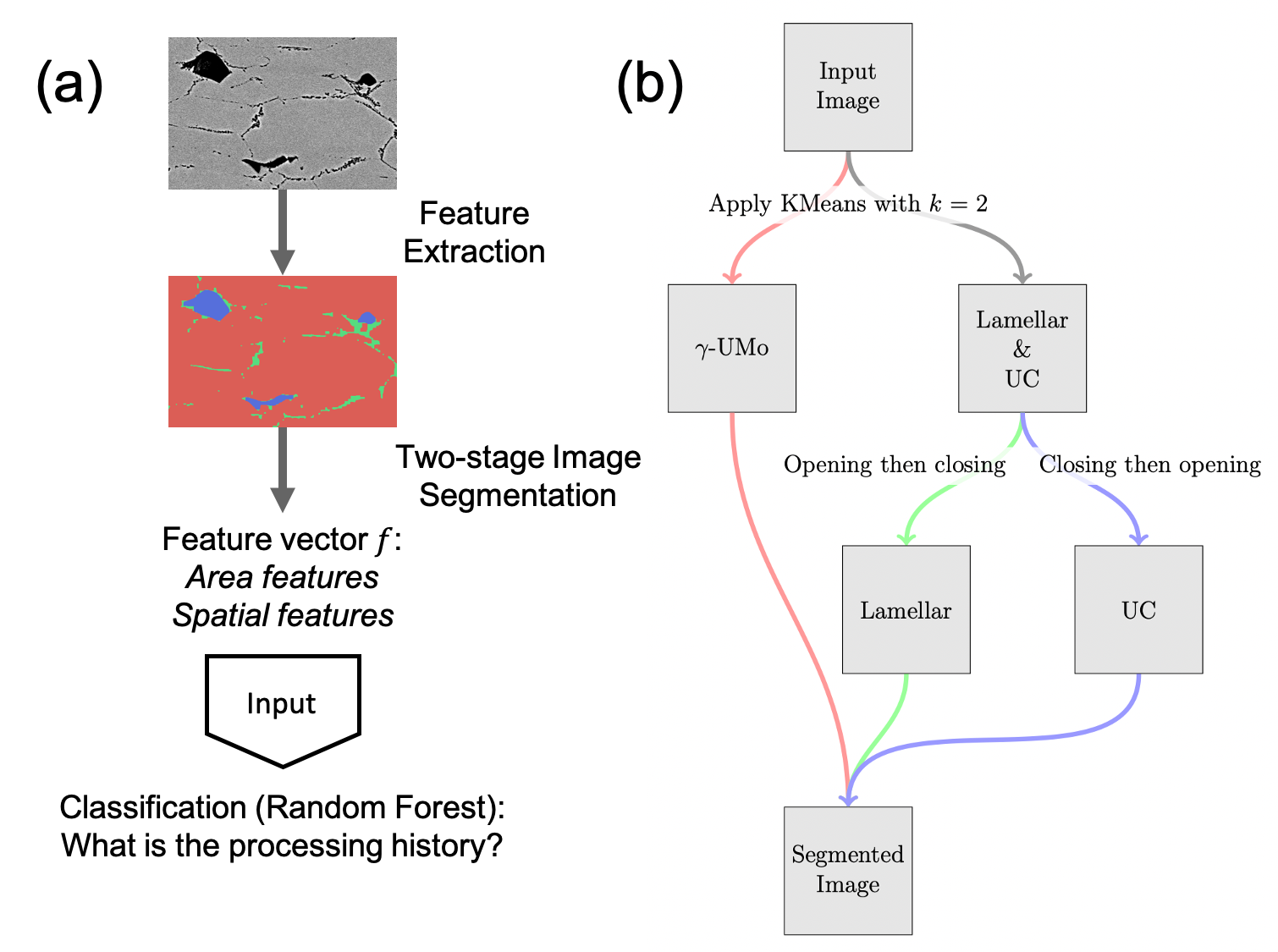}
    \caption{(a) Schematic describing the approach developed for the 10 class classification experiment performed in this work.  A two-stage segmentation approach was used, and is schematically described in (b). The two stage segmentation is described as follows: (1) stage 1 involved the use of multiple features, and (2) stage 2 utilized area-based features. A random forest model was then used for classification and five-fold cross validation was used to validate results.
    }
    \label{fig:classification-architecture}
\end{figure}

The approach to image recognition and characterization developed here is schematically described in Figure \ref{fig:classification-architecture}(a), and involves the following steps: (1) image segmentation, (2) extracting interpretable features from the image data, and (3) classifying microstructures from different classes based on extracted features.

\begin{figure*}
    \centering
    \includegraphics[scale=0.7]{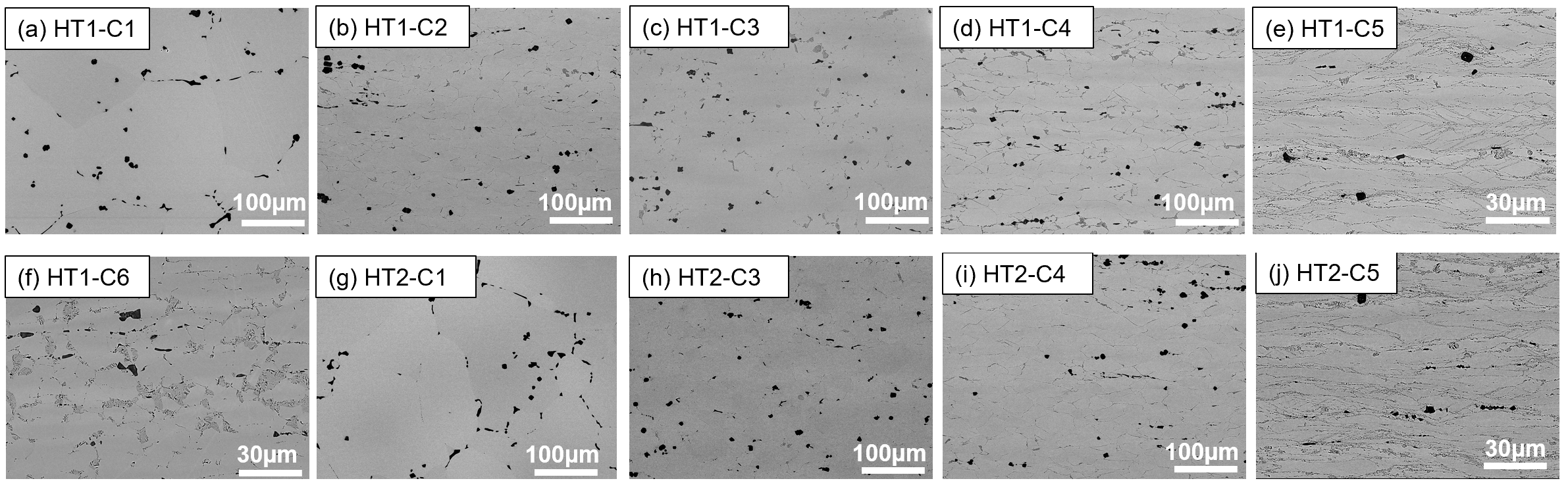}
    \caption{ Example micrographs from  different classes used and micrographs from selected class labels used in this work.  A total of ten classes were assigned as part of this effort.  In the area of classification problems this is a large number of class labels and furthermore the availability of microstructures and the subtle differences between structures further complicates the problem of successful classification.}
    \label{fig:classification-labels}
\end{figure*}

The image segmentation algorithm used here is based on our prior work where k-means was applied to classify image pixels based on the grayscale values \cite{kautz2019imagedriven}. This method is built upon the assumption that pixels correspond to different grayscale values and the differences between clusters are significant. However, in our data set not every image comes with three different phases (dictated by processing condition), which leaves us with the question whether $k$ is 2 or 3 for each image. While there are some well-known methods for choosing $k$, such as the elbow method and the silhouette method, they do not work well in our experiments. A reason for why these methods do not work here is that the grayscale shades are typically spread out evenly on a $\gamma$-compressed nonlinear scale, which means there are insignificant differences in grayscale values, even though they are noticeable to the human eye. Thus, for the image data used here, a specific $k$ value needs to be hard coded for each class. However, this hard coding requires ground truth knowledge about the image processing condition (i.e. class label). 

To overcome the limitations associated with applying k-means clustering to our image data, we developed a two-stage segmentation method which combines the k-means clustering and the image morphology. This approach is schematically described in Figure \ref{fig:classification-architecture} (b). In the first stage, we apply k-means clustering based on the grayscale values of each pixel with $k=2$. The purpose of this step is to segment the $\gamma$-UMo matrix phase from the rest of the image. In the second stage, we apply morphological opening and closing (i.e. dilation and erosion) \cite{4767941}, to remove the  fine-scale lamellae in the transformed region (so that this region is considered as a single grayscale value), and smooth the border of UC inclusions. These morphological operations aide in improving segmentation results via k-means. It is noted that for the purpose of this work, it is desirable that transformation products that appear as fine lamellae are treated as one region, where distinguishing between lamellae is not needed. 
\begin{figure}
    \centering
    \includegraphics[scale=0.35]{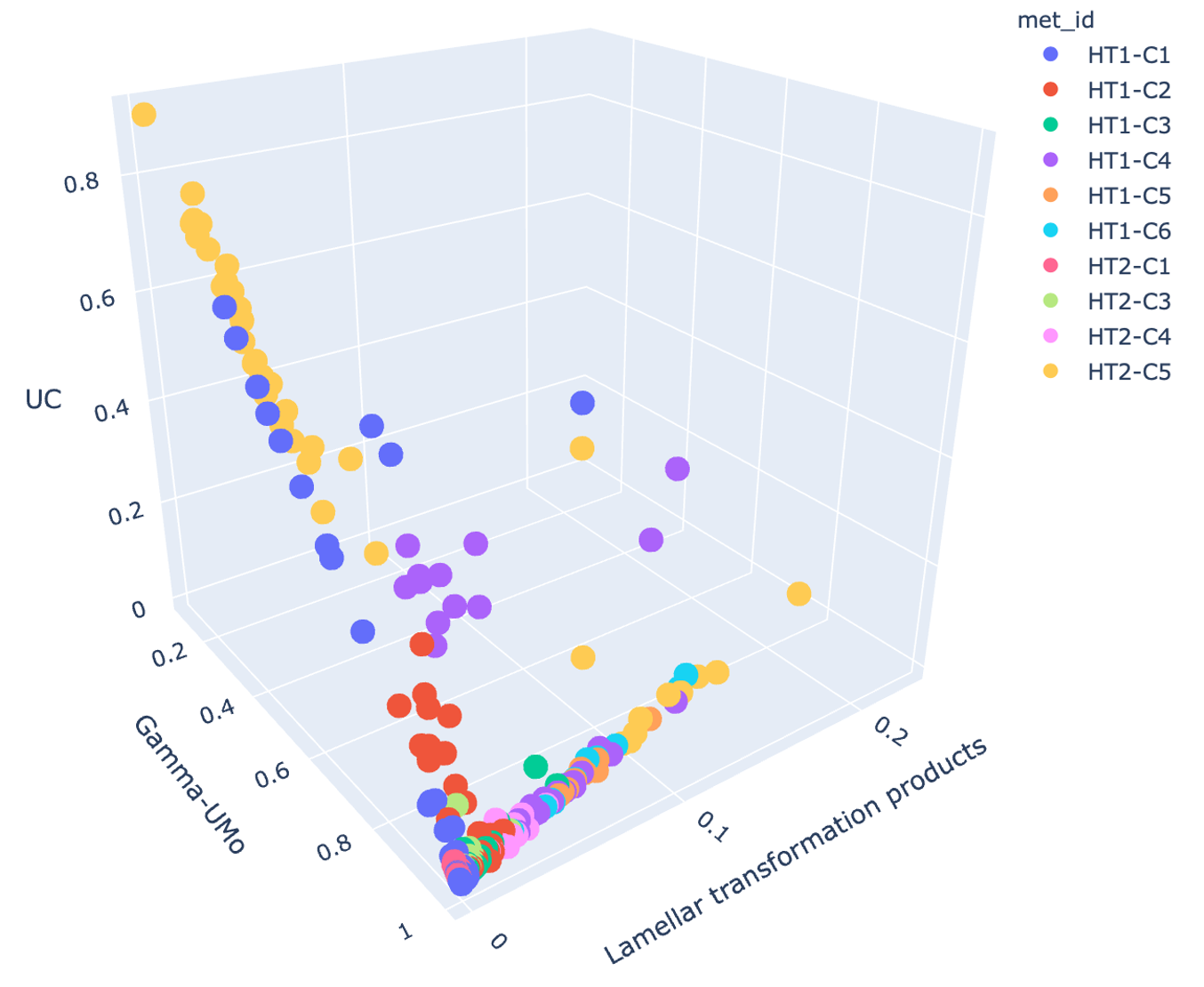}
    \caption{Visualization of area feature for each phase.}
    \label{fig:area-features}
\end{figure}

\section{Results and Discussion \label{sec:results}}

\subsection{Developing microstructure-processing relationships using discriminative learning methods}
Developing understanding of microstructure-processing relationships and improving predictive capability becomes more difficult as processing complexity increases. In our case study on the U-10Mo system, several steps are performed during fuel fabrication, and the ability to recognize what processing parameters lead to a given microstructure can allow for improved process design and quality control. However, the question of how to quantitatively describe microstructure image data in order to predict processing condition from microstructure images remains unanswered. Significant prior work has been performed in which area fraction of different phases with varying gray scale serves as a proxy for volume fraction, and is thus used as the primary quantitative microstructure descriptor \cite{Jana2017a, Jana2017, kautz2019imagedriven}. Yet, the choice of area fraction may not be the best metric when several phenomena are changing with varying processing condition, such as extent of phase transformations, distribution or fragmentation of inclusions, and change in grain size and morphology.
To measure how accurately different features can represent microstructures, we use features (area and spatial, described above) as inputs to train a Random Forest model to predict the corresponding processing history. Images are segmented and area and spatial features are extracted. In addition, we collect other texture features, such as the Haralick features and the local binary patterns (LBP), which have previously been demonstrated to represent microstructure image data well. \cite{Chowdhury2016,kautz2019imagedriven}
The following four experiments were considered to explore metrics of microstructure representation:
	\begin{enumerate}
	\item Characterization of micrographs using area features only. For the two tasks below, we train two separate Random Forest models for classification and 5-fold cross-validation is applied to evaluate the model performance:
    	\begin{enumerate}
    	    \item A 10-class classification to predict microstructure processing history (HT1-C1, HT1-C2, HT2-C1, etc.)
    	    \item A binary classification to predict the homogenization temperatures (900C-48hr or 1000C-16hr)
    	\end{enumerate}
	\item Characterization of micrographs based on area, and spatial features, in an effort to increase predictive power of our model. Similar to Experiment 1 (above), we train two Random Forest models for the two tasks listed in item 1a-b, above.
	\item Characterization of microgrpahs using area, spatial, and texture features. All features are concatenated as a single feature vector to represent a microstructure image. A model is trained to learn and predict the processing history (HT1-C1, HT1-C2, HT2-C1, etc.) of an image.
	\item Binary classification for each possible pair of processing histories (HT1-C1, HT1-C2, HT2-C1, etc.) based on area features only. This experiment provides a detailed investigation of how well area features represent micrographs.
	\end{enumerate}
Training results from these three experiments are summarized in Table \ref{tab:classification}. The model performance are measured in F1 score, defined as follows:
$$F1 = \frac{2 \cdot \text{precision} \cdot \text{recall}}{\text{precision} + \text{recall}}$$
In Experiment 1, the F1 score of the 10-class classification and the binary classification are 62.4\% and 68.5\%, respectively. In Experiment 2, spatial features are added to the area features to help improve model classification results. The performance of models is improved significantly to 78.9\% and 65.1\% for Experiment 2a and 2b, respectively. This increased performance indicates that spatial features are correlated with the processing histories. In Experiment 3, we used all the features available (both interpretable area and spatial features, and texture features), and reach an F1 score of 95.1\% for the 10-class classification task. This result serves as a benchmark for this data set and allows us to evaluate the predictive power of other models. While the area features have long been regarded as a strong indication of the microstructure processing history, from the microstructure representation experiments detailed here, we find that the predictive power of area features is actually very limited. Based on a trained Random Forest model from Experiment 3, the feature importance of the area feature corresponding to UC is 0.09, which is higher than the other 40 features. However, there are many other features from spatial and texture features with a feature importance of approximately 0.06. This conclusion can also be verified in Experiment 4, where we find that for binary classifications between two specific processing histories, the area features do not always result in high F1 scores. This finding is highlighted by very poor classification performance listed in the matrix, for example a F1 score of 61\% for the following two conditions which were both homogenized at 900C-48hr: HT1-C4 (cold rolled to 0.025 inches) and HT1-C6 (cold rolled to 0.008 inches and annealed at 700C-1hr).

\begin{table}
\caption{\label{tab:classification} Summary of experiments, features used to represent microstructure image data, metric, and performance.
 }
\begin{ruledtabular}
\begin{tabular}{|c|c|c|c|}
Experiment & Features & Metric & Performance \\ \hline
1a & Area features & F1 & 62.4\% \\
1b & Area features & F1 & 68.5\% \\
2a & Area and spatial features & F1 & 78.9\%\\
2b & Area and spatial features & F1 & 65.1\% \\
3 & All features & F1 & 95.1\% \\
4 & Area features & F1 & See Figure \ref{fig:confusion-matrix} \\ \hline
\end{tabular}
\end{ruledtabular}
\end{table}
\begin{figure}
    \centering
    \includegraphics[height=2.8in]{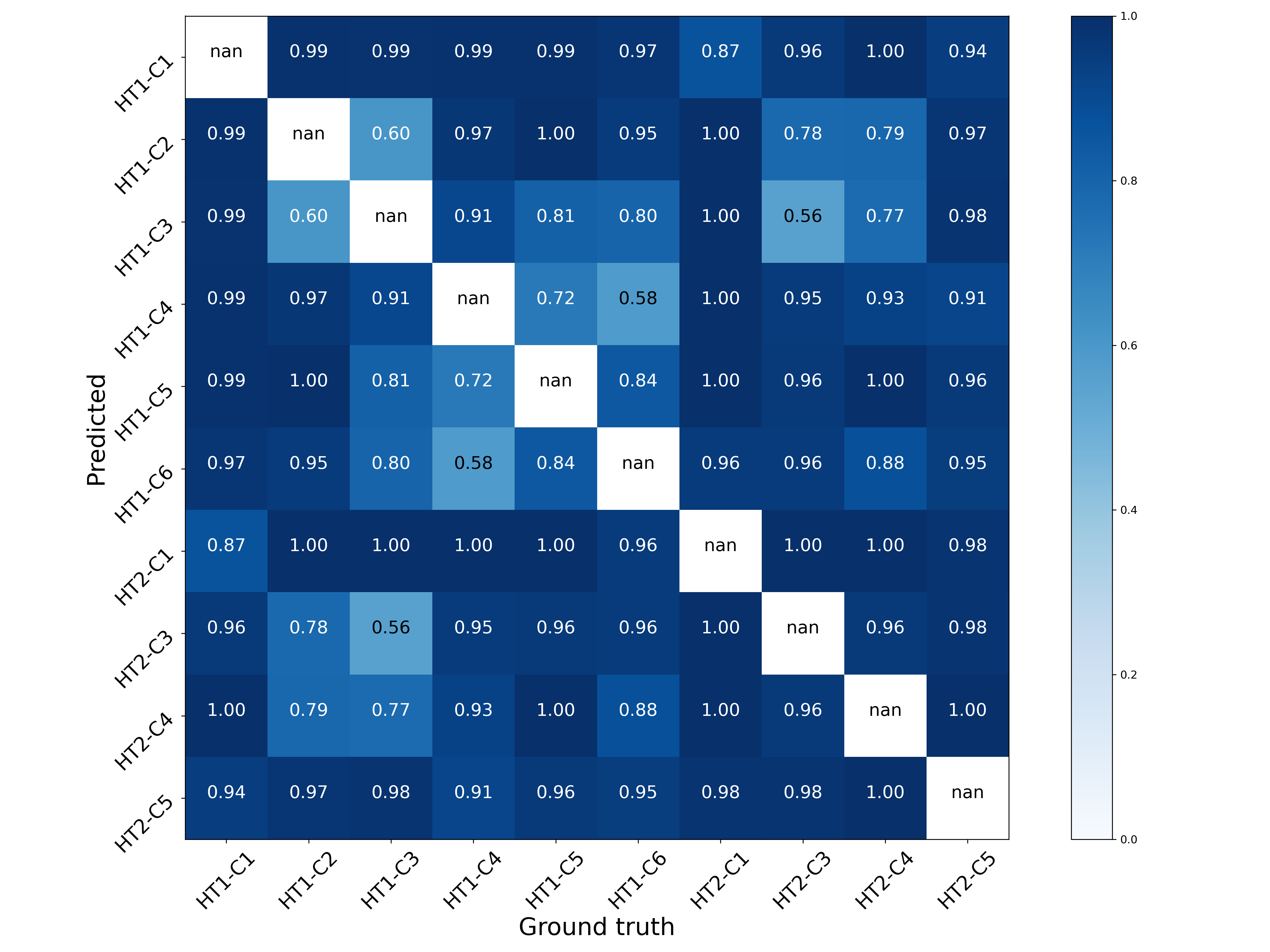}
    \caption{Model performance of binary classification for each pair of processing histories for the U-10Mo microstructure. Here, the F1 scores are reported on a 0 to 1 scale, where 1 indicates 100\% correct predictions.}
    \label{fig:confusion-matrix}
\end{figure}
\subsection{Synthetic microstructure generation using Generative Adversarial Networks}

Generative adversarial networks (GANs) have been proven successful for many image synthesis and unsupervised learning tasks \cite{pan2019recent}. It is a popular framework for representation learning, such as disentangle pose from lighting in 3D rendered images \cite{NIPS2016_6399}, and image completion, where large missing regions are synthesized utilizing the surrounding image features \cite{8578675}. Variants of GANs have surpassed many other generative models in the quality of samples as well as their underlying representation. Recently, GANs have emerged as a promising methodology for application in computational materials design, for the purpose of developing structure-property and structure-performance relations via physical simulations.\cite{Yang2018GAN, Li2018_adversarial} GANs are implemented by deep neural networks, and thus are able to capture complex microstructural characteristics. Hence, we investigate different GAN architectures here for the specific material system of U-10Mo, and the task of generating realistic artificial micrographs that could be useful in supplementing real data sets or used in an effort to predict microstructure from processing parameters.
 
A GAN framework consists of a generator, $G$, that generates samples from a noise variable, $z$, and a discriminator,$D$, that aims to distinguish between samples from the real data distribution and those from the synthetic data distribution (from the generator). The training of a GAN can be summarized as a two-player minimax game:

\begin{align*}
\min_G\max_D V(D, G) = \; & \mathbb{E}_{x\sim p_{\text{data}}(x)} \left[ \log D(x) \right] + \\
& \mathbb{E}_{z \sim p_z(z)} \left[ \log(1-D(G(z)) \right]
\end{align*}

\noindent where $p_{\text{data}}$ is the underlying distribution of real images and $p_z$ is some noise distribution.

Although the objective of the training is straightforward, the actual training can be quite unstable because of the non-convex cost functions and the high-dimensional parameter space. \cite{10.5555/3157096.3157346} In practice, the model could encounter many problems, such as vanishing gradients, where the discriminator gets too good and the generator fails to make progress, and the mode collapses (i.e. the generator collapses to a state where it always outputs the same sample). Especially in cases where we want the GAN to learn a disentangled representation of the training data and output high-quality samples, the training can be extremely hard to converge.



In this work, we make use of multiple variants of GANs to generate artificial microstructure images and demonstrate how GANs can be used to synthesize realistic images with varying resolution. In this small case study, the same set of original SEM-BSE U-10Mo micrographs described previously are used as the training set. Images are cropped into 1024 by 1024 and resized to 512 by 512 square pixels. We choose 512 by 512 as the size of training images and output samples for several reasons, including:
\begin{enumerate}
    \item High-resolution images are needed for characterization. Phases such as the lamellar transformation products may not be represented well if images are too low in resolution.
    \item The microstructural area contained within the image should be large enough to reflect the processing history. This would help to keep the variance of the training images small enough so that the GAN synthesized images represent the different classess well.
    \item The most recent GAN technology is capable of higher resolution images up to 1024 squared and in this study we wished to explore this higher resolution capability.
\end{enumerate}

Two different GAN architectures were trained and compared on the basis of artificial image quality: (1) a progressively growing GAN \cite{DBLP:journals/corr/abs-1710-10196}, where samples are generated from latent noise in original micrographs, and (2) a Pix2Pix Generative Model \cite{DBLP:journals/corr/IsolaZZE16}, where a segmentation label map is provided as input.

\subsubsection{Progressive Growing GAN}

Progressively growing GAN (pg-GAN) is an adversarial network variant that helps to stabilize the training of a high-resolution GAN. The generator is initialized with low resolution images, over which new layers are added progressively to capture finer spatial details. Each of the new layers is treated as a residual block that smoothly blends into the network when the resolution of the GAN is doubled.

Data augmentation is applied to the original set of 272 SEM-BSE images in order to increase number of images available for training. The data augmentation utilized here involves cropping original images into smaller squares with a horizontal shift, rotation by 90 degrees and horizontal/vertical flipping. Finally, images are resized to 512 by 512 square pixels. Using these methods a total of 10,880 images were available for training.

\begin{figure}
	\includegraphics[height=4in]{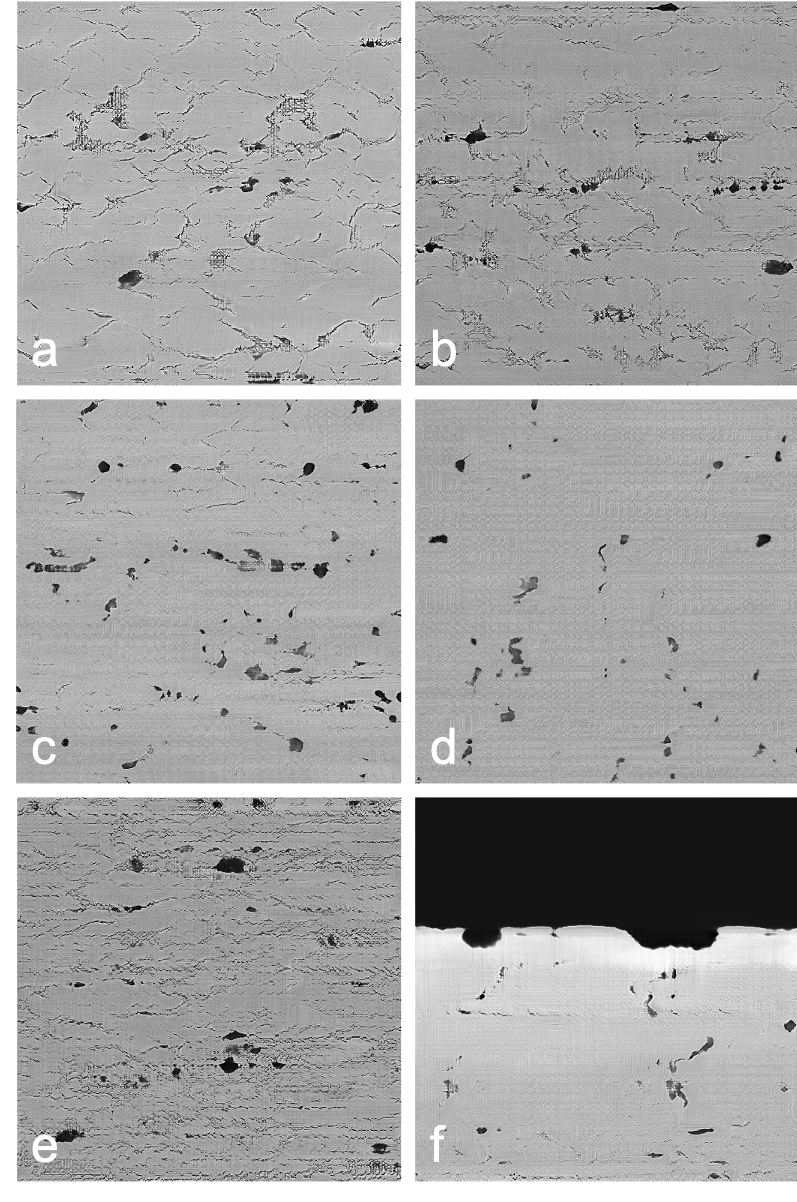}
	\caption{Example synthetic images generated by the trained Progressive Growing GAN. Images given in (a)-(f) show varying microstructural features, specifically different extent of lamellar transformation products, and distribution of carbides. The micrograph in (f) shows the edge of a fuel plate.} \label{fig:pggan-samples}
\end{figure}

We follow the model specification from the original paper \cite{DBLP:journals/corr/abs-1710-10196} with Python 3.6.9 and TensorFlow-GPU 1.13.1. We use the Adam optimizer with the default learning rate scheduling algorithm. Training images and output samples are both 512 by 512 square pixels and the training length is set to 1,000,000 images. To measure model performance, we sampled 1,000 images generated by the model, with examples given in Figure \ref{fig:pggan-samples}. The images sampled from the generated set are qualitatively close to the real data distribution although some images contain visual artifacts, such as image (f) in Figure \ref{fig:pggan-samples} (the boundary between the microstructure and the background should be a straight line). 

\begin{figure}
	\includegraphics[scale=0.24]{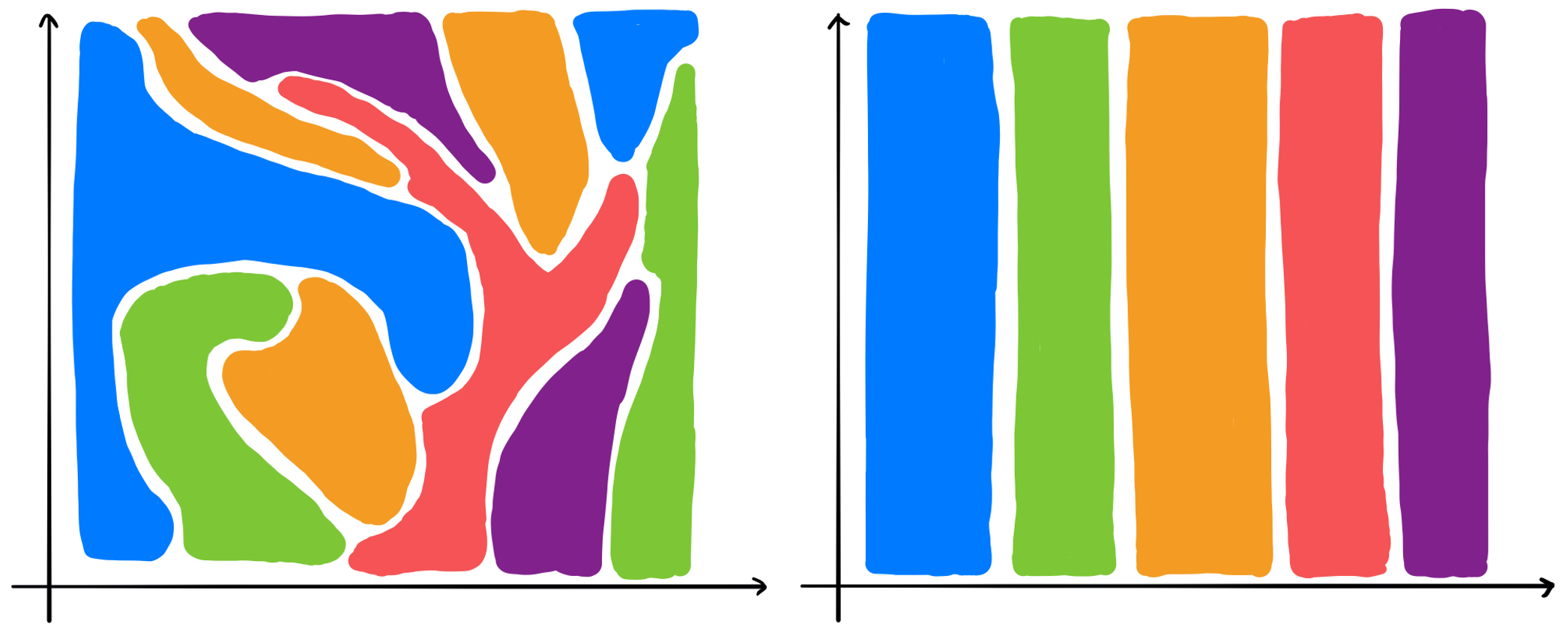}
	\caption{Schematic of an entangled representation (left) and disentangled representation (right). In the entangled representation, data are encoded in a complicated manner. In the disentangled representation, the independent variables are interpretable.} \label{fig:disentangled}
\end{figure}

Lastly, the approximated distribution is entangled, meaning that the image data are encoded in a complicated manner and the input noise variables are not interpretable (see Figure \ref{fig:disentangled}). Although the synthetic images are visually nice, we cannot interpret the role of the input noise vectors in the generation of synthetic images. This blocks us from understanding how samples from different processing histories are distributed in the learnt space of microstructure images or possibly revealing their underlying connections. Additionally, from visual examination of the Figure \ref{fig:pggan-samples} artificial images, we find some spatial patterns that are not visible in training images. Visually, we see the texture of artificial images is different, and such texture anomalies are discussed in further detail in Section \ref{sec:dft}.

\begin{figure*}
	\includegraphics[scale=0.42]{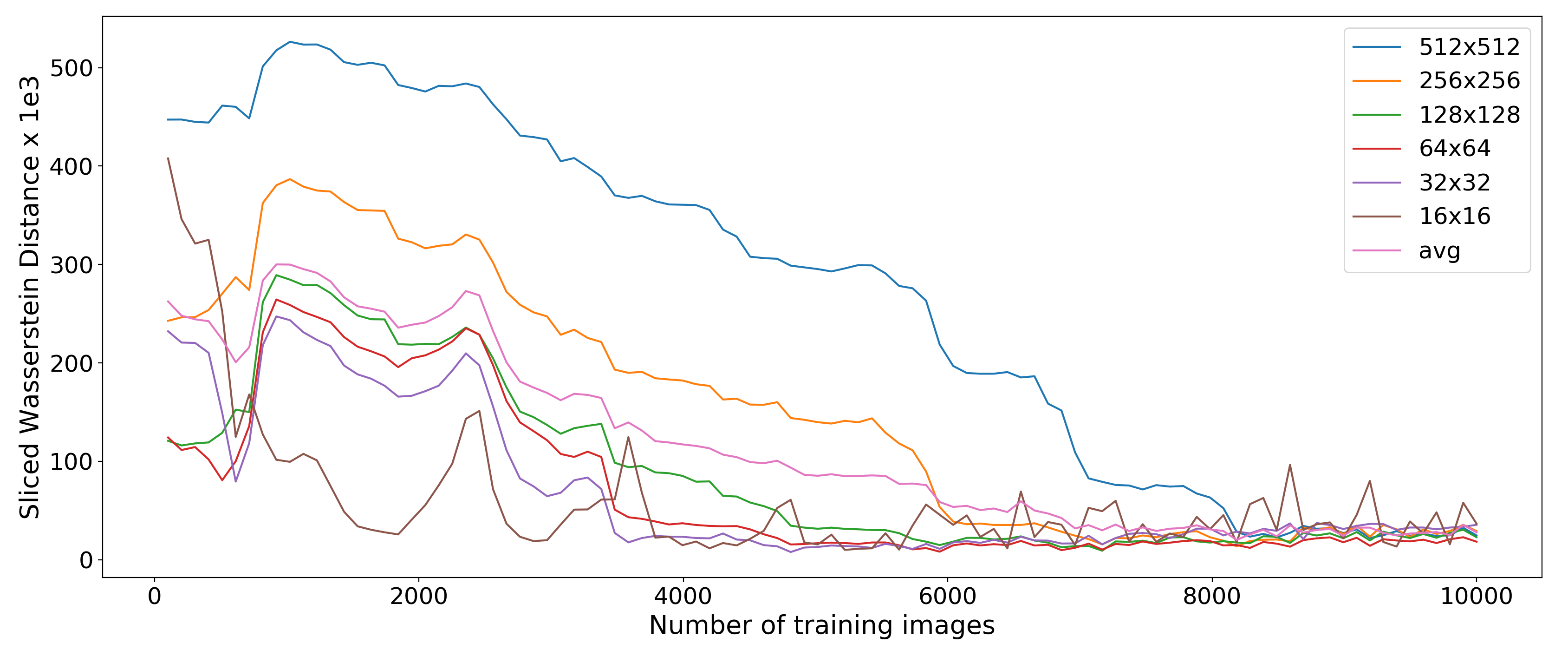}
	\caption{We use the sliced Wasserstein distance (SWD) between the training images and generated images to evaluate the model performance. This figure shows the SWD under different resolutions as the training progress.} \label{fig:pggan-swd}
\end{figure*}

To better assess the GAN results, we turn to automated methods such as the sliced Wasserstein distance (SWD) \cite{Bonneel_2014} to measure how similar artificial images are to the training set over different scales. We measure the SWD with the checkpoint images during training and plot the distances with respect to the number of training images fed into the model, with results given in Figure \ref{fig:pggan-swd}. We find that the SWD at different resolutions generally decreases as the training proceeds. The model converges after approximately 8,000 thousand training images. Even with limited micrographs for training, the progressive growing GAN can still learn the real data distribution well, as demonstrated by the realistic synthetic images shown in Figure \ref{fig:pggan-samples}(a)-(f).

\subsubsection{Pix2Pix Generative Model}

Unlike the Progressive Growing GAN, Pix2Pix GAN is a conditional GAN \cite{DBLP:journals/corr/MirzaO14} variant that learns a mapping from some extra given information $y$ (``condition'') and input noise $z$ to output images. The objective of a conditional GAN is given by:

\begin{align*}
\min_G\max_D V(D, G) = \; &  \mathbb{E}_{x \sim p_{\text{data}}(x)} \left[ \log D(x \mid y) \right] + \\
& \mathbb{E}_{z \sim p_z(z)}\left[ \log (1-D(G(z\mid y))) \right]
\end{align*}

\begin{figure}
	\includegraphics[height=3in]{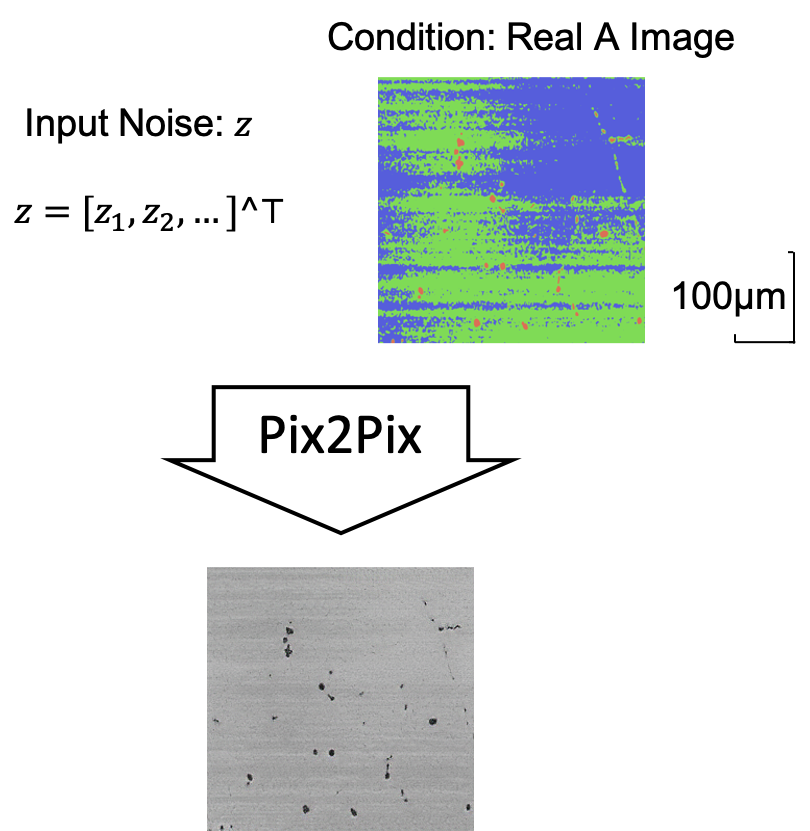}
	\caption{Pix2Pix Generative Model takes a label image as input and outputs a synthetic microstructure image. In this example, the Real A image is generated from a sample with the HT1-C1 processing history.} \label{fig:pix2pix}
\end{figure}

The Pix2Pix model is widely used for image-to-image translation, such as image style transfer, labels to Facade, and edges to photo. In this work, we use the labels generated from the segmentation algorithm detailed above as style A and realistic microstructure images as style B. Overall, the Pix2Pix generative model takes a labeled image as input and generates a realistic microstructure image, as shown in Figure \ref{fig:pix2pix}. We note here that the segmented image given as the input includes some noise (due to charging from the sample in the SEM) that was segmented as a separate phase. Although from a segmentation point of view, this is not an accurate representation of the microstructure (i.e. the noise is not an important microstructure feature of the image), it does mean that the charging artefact in the original image is accurately captured in the synthetic image, thereby making the synthetic image realistic when compared to the original image data.
The same set of 272 original images is used, after removing the scale bar and cropping each original image into squares. These images are then used for training, and are referred to as real B, which serve as the ground-truth images (style B). Prior to training, real image labels are prepared, where the real images are referred to as real A (the model input). Image segmentation, such as the algorithm suggested in \cite{kautz2019imagedriven} or the one described here in Section \ref{sec:methods_2a}, can be applied so that for each image in real B, we have a label image in real A.

\begin{figure}
	\includegraphics[height=2in]{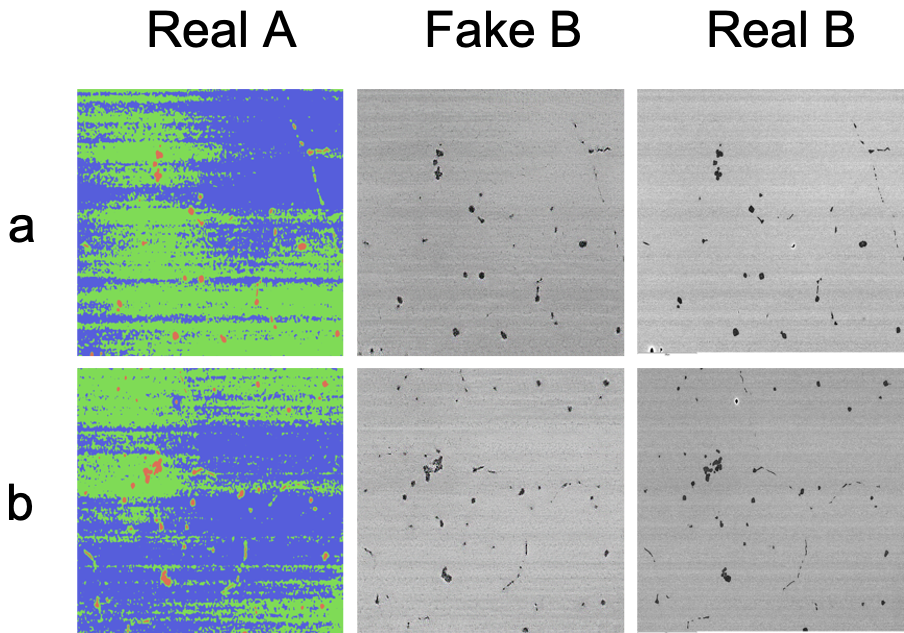}
	\caption{Randomly sampled images from the trained Pix2Pix model. Each row is a different sample and from left to right are the images from real A, fake B (model output), real B (ground-truth).} \label{fig:pix2pix-samples}
\end{figure}

We use the default model specification described in the Pix2Pix model paper \cite{DBLP:journals/corr/IsolaZZE16}. After the training is finished, 50 synthetic images are randomly sampled and some of them are displayed in Figure \ref{fig:pix2pix-samples}. From the sampled images, we can tell that the synthetic images generated by the Pix2Pix model are visually close to the ground-truth. With sufficient information from the label images, they are more realistic than those sampled from the Progressive Growing GAN. While spatial patterns are not visible in these sampled images, we apply the same measurements as we did on the Progressive Growing GAN as a comparison, which can be found in Section \ref{sec:dft}.

While the outputs from the Pix2Pix model are visually more realistic, they require additional information as inputs. They ignore the distribution of microstructures in the image and focus on the learning and simulating the textures in real images. Training a high-resolution GAN with interpretable conditions remains a challenge. For future studies, training a high-resolution GAN could be split into two steps in which we first focus on synthesizing the underlying representation of the microstructures (such as the label image), and then in a second step, adding texture to the image.

\subsubsection{Analysis of Microstructure Distribution Learnt by the pg-GAN \label{sec:micro-dist-analysis}}

In this section, we measure the differences between the microstructure represented by the real images and synthetic images. We apply the characterization pipeline introduced in Section \ref{sec:methods_2}. Considering the visual differences (image ``sharpness'' and local patterns) between real images and synthetic images, and that original images are resized to a smaller size before being used for GAN training, we use a slightly different parameter setting (given in Table \ref{tab:params2}) from the previously described to ensure the quality of image segmentation.

\begin{table}
\caption{\label{tab:params2} Parameters selected for microstructure characterization of synthetic images.}
\begin{ruledtabular}
\begin{tabular}{|r|c|c|}
&\mbox{\textbf{Parameter}}&\mbox{\textbf{Value}}\\ \hline

\textbf{Preprocessing} & Sharpen & (3, 3) \\
    & Noise-reducing &	Bilateral \\
    &	diameter	&	15	\\
	&	sigma color	&	75	\\
	&	sigma space	&	75	\\	\hline
\textbf{Segmentation} & k-means & kmeans++ center initialization \cite{10.5555/1283383.1283494} \\
	&	Lamellar	&	(7,7) closing then (7,7) opening	\\
	&	UC	&	(5,5) opening then (3,3) closing	\\ \hline
\end{tabular}
\end{ruledtabular}
\end{table}

\begin{figure*}
	\includegraphics[height=3in]{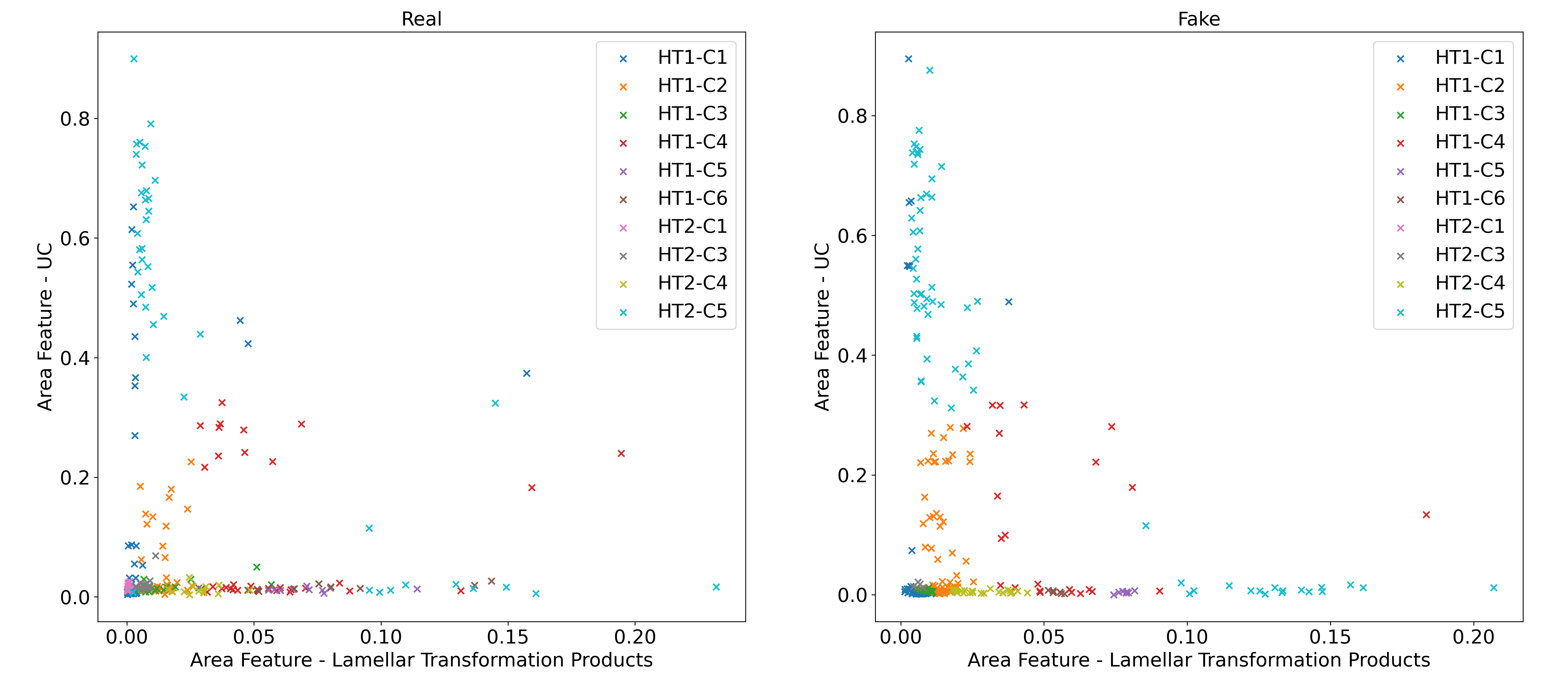}
	\caption{Comparison between area features from real images and synthetic images. The processing histories of the synthetic images are predicted by a Random Forest model trained on features from real images.}
	\label{Fig:synthetic-area-features}
\end{figure*}

We generate 300 synthetic images by random sampling from the trained pg-GAN model. Synthetic images are segmented and area features, spatial features, and texture features are collected. The processing histories of the synthetic images are predicted by the trained model from Experiment 4 in Section \ref{sec:results}. The area features of Lamellar Transformation Products and UC are plotted in Figure \ref{Fig:synthetic-area-features}, along with the predicted processing histories. It can be found that the area features collected from real images and area features collected from synthetic images come from similar distributions.

To quantitatively measure whether the synthetic image are from the same distribution as the real images, or how well the pg-GAN has learnt to represent the microstructure, we carry out the two experiments below.


From the previous experiments, we have collected area features, spatial features, and texture features from the 272 original images and the 300 synthetic images. Two models are trained to classify whether a specific features is collected from a real image or a synthetic image. The first model uses the area features as the only input, while the second model takes all the features (area, spatial, texture) as the input. Both models use the Gaussian Process Classifier (GPC) to learn two distributions for the features from real images and synthetic images. With 5-fold  cross-validation, the accuracies for the two models are 52.6\% and 52.4\% respectively. The fact that both models fail to distinguish between features collected from real images and synthetic images supports our assumption that the pg-GAN managed to learn the underlying distribution of microstructure well.

\begin{figure}
	\includegraphics[height=2.8in]{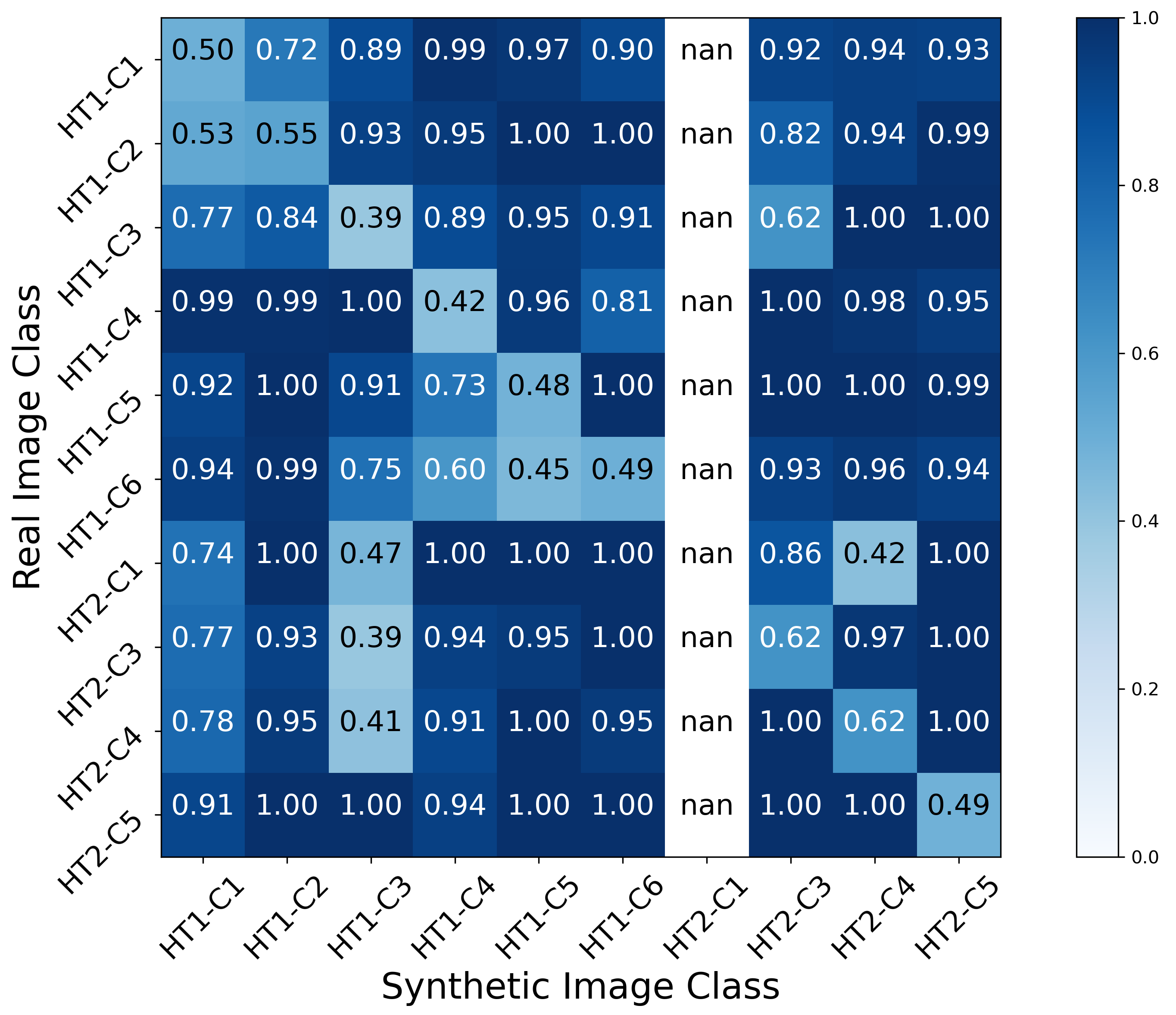}
	\caption{Results of binary classification between area features collected from real class A images and area features collected from synthetic class B images. Model performance are measure in F1 scores with 5-fold cross-validation. Notice that there are no synthetic images predicted as HT2-C1 in the 300 synthetic images.}
	\label{Fig:synthetic-cm}
\end{figure}

Here, we consider the area features collected from the 272 original images and the 300 synthetic images. For each pair of processing histories (e.g. HT1-C1 and HT1-C2), we train a binary classification model to classify between the area features collected from real HT1-C1 images and the area features collected from synthetic HT1-C2 images. Again, the Gaussian Process Classifier (GPC) is used as the model classifier and the classification results are reported in Figure \ref{Fig:synthetic-cm} as F1 scores with 5-fold cross-validation. It is noted here that since there are no synthetic images predicted as HT2-C1 in the 300 synthetic images, the column corresponding to HT2-C1 is left empty in the matrix.

It can be found that on the main diagonal matrix, where the real area features and synthetic area features are from the same processing histories, the F1 scores are generally low. This finding is consistent with our assumption that the pg-GAN managed to learn the underlying distribution of microstructure well. Also, the classification performance in other cells off the main diagonal are relatively high with a few exceptions. For those off-diagonal cells with low F1 scores, such as the classification between HT2-C3 and HT1-C3, they are quite consistent with the classification results reported in Figure \ref{fig:confusion-matrix}. The poor classification performance for this particular example of HT2-C3 and HT1-C3 may be attributed to the very similar microstructure generated after few processing steps, where the only difference between these two conditions is the homogenization temperature and time.

By evaluating classification performance between features from real and synthetic images, and between area features for different processing histories, we can compare the area features, spatial features, and texture features collected from real images and synthetic images. In previous experiments, we have shown that these features are quite good at characterizing the microstructure. Especially for the Random Forest model that takes all features as the input, processing history of microstucture can be predicted with an F1 score of 95.1\%. However, these features are not explicitly considered during the GAN training. From the fact that features collected from real or synthetic images are not distinguishable indicates that the pg-GAN model managed to learn the underlying distribution of microstructure well. Although the results from binary classification are highly based on the processing histories that are predicted by a Random Forest model, and not originally given by the pg-GAN, the results can serve as a comparison with the experiment results in Section \ref{sec:results} and suggest the underlying representation learnt by the pg-GAN is similar to the microstructure representation given by the real images.

\subsubsection{Texture Differences in Real Versus Synthetic Micrographs \label{sec:dft}}

Sub-regions of images from real microstructures and images generated by the GAN were subjected to the Discrete Fourier Transform (DFT) operation. This was performed with the objective of studying the spatial patterns exhibited by these two classes of images. 

DFT samples a discrete set of frequencies corresponding to the size of the  image in the spatial domain. In the Fourier domain, the intensity at frequency point (k,l) is calculated by:

\begin{equation}
F(k,l) = \sum_{i=0}^{N-1}\sum_{j=0}^{N-1} f(i,j)e^{-\frac{2\pi}{N}(ki + lj)}
\end{equation}

\noindent where f(i,j) is the pixel intensity at position (i,j) in the real space.
Images exhibiting geometric or spatial patterns tend to amplify specific frequencies in the Fourier domain, and hence the Fourier transform of an image can be used to highlight spatial patterns present in the image. For example, when an image exhibits horizontal patterns seperated by a pixel distance of WIDTH/2, its Fourier Transform exhibits a local maxima at (k,l) = (2,0).

\begin{figure}
	\includegraphics[scale=0.7]{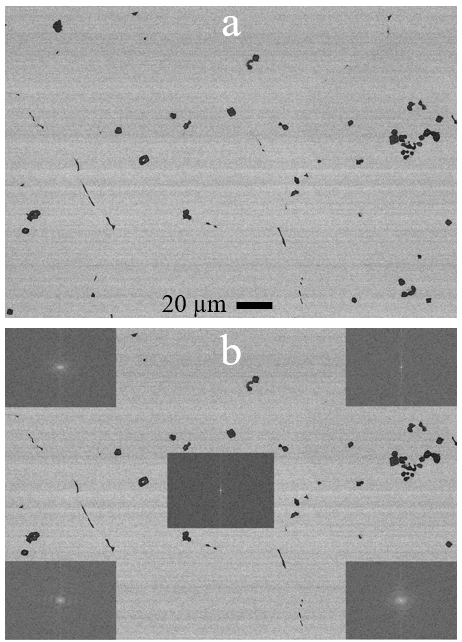}
	\caption{(a.) A characteristic image of a real microstructure, showing visible texture in the horizontal direction. (b.) Fourier Transforms of different sub-regions of size 200 x 200, superimposed upon the original image at the corresponding region. The background pattern is reflected in the thin vertical lines on the transform plots.} \label{Fig:real-ft}
\end{figure}

\begin{figure}
	\includegraphics[scale=0.8]{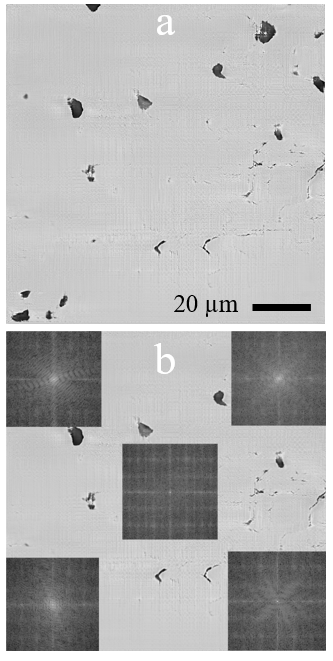}
	\caption{(a.) A characteristic image of a microstructure generated by pg-GAN. (b.) Fourier Transforms of different sub-regions of size 200x200, superimposed upon the original image at the corresponding region. Though these synthetic images show strong background features, the presence of these features does not bias the discriminator. } \label{Fig:fake-ft}
\end{figure}

In this work, DFT was implemented in Python using the OpenCV \cite{opencv_library} and Numpy \cite{Oliphant2006} libraries. The `dft' method in OpenCV was used to perform the transform, and the `fftshift' method in Numpy was used to shift the zero-frequency (DC frequency) to the center of the transform. The insets on Figures \ref{Fig:real-ft} and \ref{Fig:fake-ft} are magnitudes of the transforms that have been subjected to a log filter, in order to visualize the local maximas effectively.

Figure \ref{Fig:real-ft} shows a characteristic example of an image of a real microstructure and the Fourier transforms performed on the different sub-regions of the image. The sub-regions selected for this analysis are all 200px $\times$ 200px in size. It can be observed that the horizontal texture manifests itself in thin vertical frequency lines on the transforms.

Figure \ref{Fig:fake-ft} shows a characteristic example of a synthetic image generated from an adversarial network and the Fourier transforms performed on the different sub-regions of the image. It can be seen that the synthetic images are characterized by strong patterns, which manifest as local maximas at several frequency points on the transformed images. The Fourier transforms of a synthetic image exhibit more local maximas than the corresponding transforms of a typical image of a real microstructure.

\begin{figure*}
	\includegraphics[height=6cm]{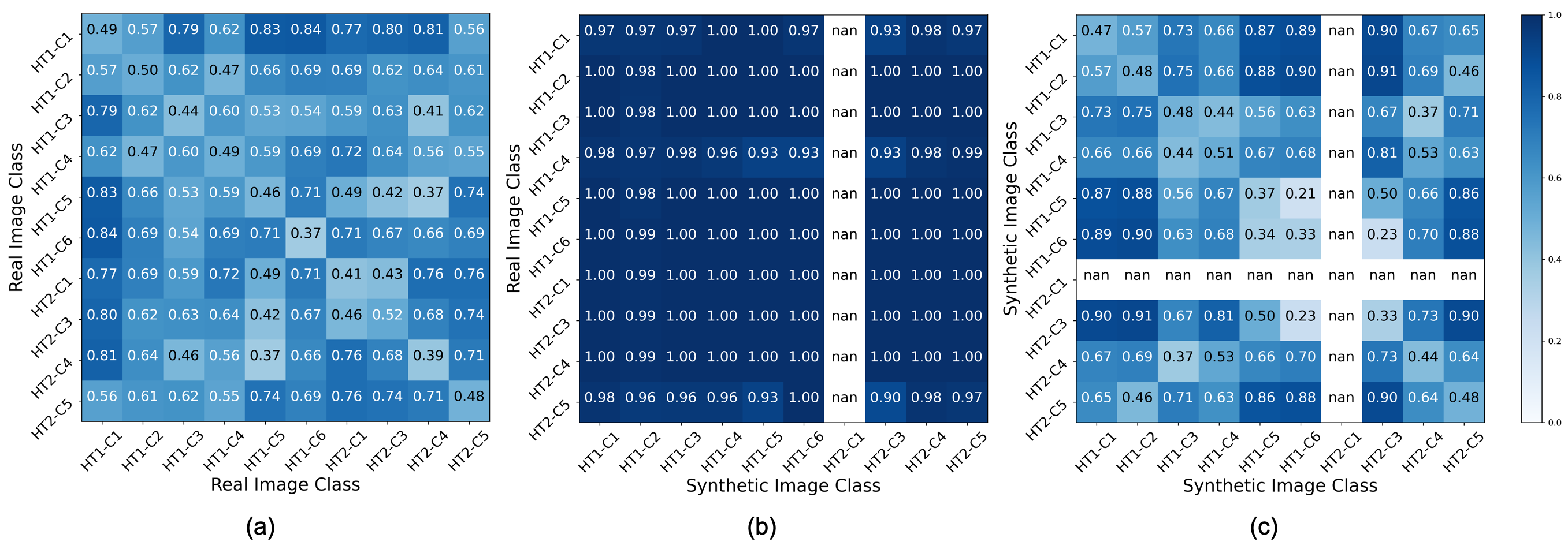}
	\caption{Binary classification between microstructure images from different processing histories based on DFT features. The model performance are measured in F1 scores. From left to right, the matrices measure the model performance between: (a) real images and real images, (b) real images and synthetic images, and (c) synthetic images and synthetic images.} \label{fig:dft-cm}
\end{figure*}

In order to quantitatively analyze how the magnitudes of Discrete Fourier Transform can be used to characterize the texture in real and synthetic images, as well as microstructure from different processing histories. We collect the magnitudes of the transform as a feature vector, referred to as DFT features. More specifically, after the DFT process, we take the magnitudes from the 9x9 region in the center and flatten them into a 1-dimensional feature vector. We collect the DFT features from the 272 real images and the 300 synthetic images, which are prepared in Section \ref{sec:micro-dist-analysis}. Two experiments are carried out:

\begin{enumerate}

\item Using the DFT features only, we train a Gaussian Process Classifier model to predict whether a DFT feature vector is collected from real microstructure images or synthetic microstructure images. With 5-fold cross-validation, the accuracy of the model is 99.4\%.

\item For each pair of processing histories (e.g. HT1-C1 and HT1-C2), we train a Gaussian Process Classifier model to predict if a DFT feature vector is collected from real HT1-C1 images and real HT1-C2 images. Similar experiments are conducted between real HT1-C1 images and synthetic HT1-C2 images, and between synthetic HT1-C1 images and synthetic HT1-C2 images. Model performance are measured in F1 scores, as reported in Figure \ref{fig:dft-cm}.

\end{enumerate}

It can be concluded that as a feature vector, the DFT magnitudes can discriminate between real and synthetic images well, but fail to characterize microstructure from different processing histories. Considering the synthetic images generated from the pg-GAN model, the DFT features are good indicators that the synthetic images are not perfect.

It should be noted that the synthetic image used for the analysis above was generated after the adversarial network had reached a steady state (i.e., at a point when the network can no longer distinguish between a real and a synthetic image). Thus, it is reasonable to conclude that though the images generated by GANs exhibit spatial patterns that are not present in the training images, the presence of these spurious patterns is not significant enough to bias the discriminator. The likelihood of the occurrence of such patterns must be taken into consideration for classification problems such as texture detection in a microstructure dataset.

\section{Challenges and Best Practices}
Several challenges associated with recognition and quantification of microstructure image data exist, due to various limitations inherent in materials science studies. The two major challenges are limited size of datasets and imbalanced datasets. Many machine learning (particularly deep learning) models require large datasets for training. A deep learning model generalizes to test datasets better as the size of training set is increased. In a typical environment for microstructure imaging, generating high quality micrographs in large numbers is dependent upon metallographic sample preparation, microscope operator skill, facilities, and time. For objectives such as multi-class classification, training from real data sets could lead to a situation where few classes have a disproportionate number of images, highlighted in prior work\cite{Baskaran2020}, and in the U-10Mo data set studied here. Imbalanced datasets can result in a reduced classification accuracy for the class with the disproportionately lower number of images, even if the overall accuracy is within acceptable tolerance. With respect to the case study of GAN seen earlier in the study, an imbalanced dataset may also result in a reduced significance for the spatial patterns present in the disproportionate class among the patterns exhibited by the synthetically generated images. Given these challenges, the authors suggest several best practices, in addition to the current work, to produce meaningful results from an image driven machine learning approach to microstructure recognition and quantification: 

\begin{enumerate}
	\item Shallow learning and conventional learning techniques: Convolutional Neural Networks and other algorithms such as SVMs or ensemble classifiers are capable of performance within an acceptable tolerance for simple classification problems.
	\item Semantic Segmentation: DeCost et al\cite{DeCost2018} demonstrated the use of the semantic segmentation algorithm to isolate objects in the microstructure dataset
	\item Serialization of techniques through a task pipeline: Prior work  has demonstrated a method for quantitative feature extraction by automating the algorithm selection process through a task pipeline.\cite{Baskaran2020}
	\item Dataset augmentation: The user may artificially populate the training dataset by methods such as cropping, rotating, and adding uncorrelated noise to the original images in order to ensure that the training process generalizes reasonably well to test datasets.
	\item Automated algorithm selection and hyperparameter optimization in machine learning: Given the size and scale of datasets in material science, automatic methods for selection of algorithms and hyperparameters \cite{chowdhury2019quantifying, hutter2019automated, kotthoff2017auto, Bengio2011, bergstra2011algorithms, li2017hyperband} can prove to be a viable option for fine tuning the parameters of algorithms and improve generalization performance as much as possible given the scarcity of data to learn from.
\end{enumerate}

\section{Conclusions}

In this work, we perform multi-class classification for the purpose of linking microstructure to processing condition. The original data set consists of micrographs for ten different thermo-mechanical processing conditions of a U-10Mo alloy. We evaluate the classification model performance for different microstructure reperesentations, and results reveal that area, spatial, and texture information are needed for accurately describing image data. Using this newly developed microstructure representation, an F1 score of 95.1\% was achieved for distinguishing between micrographs corresponding to ten different thermo-mechanical material processing conditions. Generative adversarial networks were also explored to better understand if synthetic image data could be used to supplement small, imbalanced original image data sets. Two different networks were trained and tested to assess performance: Progressive Growing GAN and Pix2Pix GAN. We find that the Progressive Growing GAN introduces spatial patterns that are not present in original image data. Texture detection in a microstructure dataset, might be adversely affected by the presence of such spurious patterns. Our work highlights that semantically meaningful segmentation alone may be insufficient in representing image data, particularly as the complexity of material processing and resultant microstructure increase. Hence, the need for predictive or generative methods is a frontier in materials science and engineering, and should be leveraged in future studies to accelerate the materials design and characterization process. 

\section*{Data Availability}
The data that support the findings of this study are available from the corresponding author upon reasonable request.
\section*{Author Contributions}
W.M. performed all machine learning methods development. E.K. generated a portion of image data, assisted in interpretation of results, and lead manuscript writing with W.M.. V.J. contributed to discussions of results and significance of microstructure evolution in the U-Mo system. A.B. analyzed synthetic image data provided by W.M. A.C. contributed to discussion of machine learning methods and results. B.Y. and D.L. conceptualized and directed research performed. All authors contributed to manuscript preparation.

\begin{acknowledgments}
Experimental and data analysis work was conducted at Rensselaer Polytechnic Institute, and Pacific Northwest National Laboratory operated by Battelle for the United States Department of Energy under contract DE-AC05-76RL01830. A portion of this work was funded by the U.S. Department of Energy National Nuclear Security Administration's Office of Material Management and Minimization. The authors also wish to acknowledge personnel at PNNL for experimental work done to generate data used in this work, in particular the following individuals: Shelly Carlson, Mark Rhodes, and Jesse Lang (PNNL) for metallographic sample preparation, and Alan Schemer-Kohrn (PNNL) for expertise in electron microscopy and image acquisition of image data used in this work.
\end{acknowledgments}



\section*{References}
\bibliography{aipsamp}

\end{document}